# Размерная зависимость теплофизических свойств наноразмерных тел в представлении нанотермодинамики Хилла

Рассматривается задача построения размерной зависимости для теплофизических свойств наноразмерных тел на примере удельной теплоёмкости. Методологической основой является нанотермодинамика Хилла. Вынужденный отказ от постулата аддитивности приводит к задаче построения обобщённой статистической суммы на основе комбинаторно-статистического усреднения. Раздел «Статсумма без аддитивности» описывает построение такой суммы. Обобщённая статистическая сумма включает объёмно-зависимый множитель, для которого строится статистическое приближение. Следствием полученного результата является выражение для размерной зависимости термодинамического потенциала на примере свободной энергии Гиббса (раздел «Термодинамический потенциал»). Сочетание понятия размерного множителя (раздел «Размерный множитель») и выражений для теплоёмкости позволяет построить размерную зависимость для удельной теплоёмкости наноразмерной частицы (раздел «Размерная зависимость теплоёмкости»).

Ключевые слова: наночастицы, нанотермодинамика, размерная зависимость, теплофизика, теплоёмкость

## Проблема

Одной из термодинамических методологий, приложимых к наноразмерным объектам, является нанотермодинамика Хилла [1]. Особого рода ансамбль, вводимый в этой методологии, будем здесь называть *хилловским коллективом*. Поскольку хилловский коллектив макроскопичен и по определению состоит из невзаимодействующих между собой *малых (хилловских) объектов*, для него справедливо основное соотношение статистической термодинамики между одним из термодинамических потенциалов и логарифмом статистической суммы [1]:

$$\Phi = -k_B T \ln \Xi_K = -k_B T \ln \Xi^{\mathbf{K}},$$

где $\Phi$ — потенциал; $\Xi_K$ — статистическая сумма <u>для коллектива в целом</u>; $\Xi$ — статистическая сумма <u>для отдельного объекта</u>; $\mathbf{K}$ — количество объектов в коллективе.

При этом статистическая сумма в методологии Хилла используется в обобщённом (по Хиллу) виде, то есть, уточнена добавлением множителей—статистических весов, зависящих от внутренних параметров. Так, например, обобщённая каноническая статистическая сумма в условиях NVT [1, ч.1, §1-3, 1-4;

1, ч.2]:

$$\Xi_{NVT} = \sum_q \exp[-\beta \mathcal{E}_q(N,V)],$$

где $\mathcal{E}_q(N,V)$ — энергия квантового уровня $q$, формально зависящая от экстенсивных переменных $N$ и $V$; $N$ — количество структурных единиц (атомов, ионов, молекул) в объекте; $V$ — объём объекта; $\beta = (k_B T)^{-1}$ — температурный параметр.

Обобщённая статистическая сумма для хилловского объекта в условиях NPT [1]:

$$\Xi_{NPT} = \sum_q \sum_\nu \exp(-\beta \mathcal{E}_q) \exp(-\beta p \mathrm{v}_\nu) = \sum_\nu \exp(-\beta p \mathrm{v}_\nu) Z, \qquad (1)$$

где $q$ — индекс энергетических состояний объекта; $\mathcal{E}_q$ — энергия состояния с индексом $q$; $\nu$ — индекс различных величин объёма объекта; $\mathrm{v}_\nu$ — объём объекта, обозначенный индексом $\nu$; $Z$ — каноническая статистическая сумма для целого объекта.

Далее — в традиционном подходе — полная статсумма (факторизуемая) разлагается на элементарные множители [2]:

$$\Xi_{\text{факт}} = \sum_\nu \exp(-\beta p \mathrm{v}_\nu) Z = \sum_\nu \exp(-\beta p \mathrm{v}_\nu) z^L, \qquad (2)$$

где $z$ — каноническая статистическая сумма для одного элемента (например, локальной моды); $L$ — количество элементов; .

Однако *предположение об аддитивности* [2], разрешающее такое разложение, в наноразмере оказывается некорректным, так что выражение (2) использовать нельзя, по крайней мере, без существенных дополнений.

## Статсумма без аддитивности

Построить выражение для статистической суммы без предположения об аддитивности можно либо для объекта в целом, вводя необходимые упрощения (например, на основе *вибронного метода* [3] и комбинаторной статистики

[4]), либо получив некое приближение, приводящее неаддитивные эффекты к одному структурному элементу тела, что позволяет воспользоваться простым в обработке (аналитической, численной) выражением (2) [4]. Во втором случае, который и рассматривается здесь, основной проблемой оказывается расчёт объёмо-зависимого множителя в (1).

Вообще вычисление объёма наноразмерных объектов является трудной задачей в том же смысле, что и определение (измерение) их размеров [5]. Можно обойти эту трудность, заметив, что объект не может быть бо́льшим, чем его внешняя поверхность, образованная некоторой долей структурных единиц (атомов, ионов, молекул) целого объекта. Далее можно заменить в выражении (1) суммирование по «объёмным состояниям» на суммирование по всем возможным сочетаниям координат всех структурных элементов внешней поверхности тела. Структурными элементами здесь называются те сущности, движение (изменение) в которых учитывается в статистической сумме; обычно — квантовые осцилляторы. Возможность такого решения вообще обусловлена либо ограниченностью набора индивидуальных состояний для структурных элементов, либо возможностью получить аналитическое выражение в случае бесконечного количества таких состояний.

Будем считать, что внешняя поверхность топологически двумерна без особенностей. Тогда количество структурных элементов в ней можно выразить некой квадратичной формой [6, 7]

$$P_2(x_1, x_2, x_3) = \mathbf{O}(2x_1 x_2) + \mathbf{O}(2x_1 x_3) + \mathbf{O}(2x_2 x_3) + \\ + \mathbf{o}(x_1 x_2) + \mathbf{o}(x_1 x_3) + \mathbf{o}(x_2 x_3)  \qquad (3)$$

где $x_k$, $k = 1, 2, 3$ — линейный размер в *кристаллографическом базисе*, выраженный в безразмерных единицах (периодах кристаллической решётки); $\mathbf{O}(...)$, $\mathbf{o}(...)$ — стандартная *O-нотация*.

Для упрощения вида выражений будем считать, что все векторы базиса попарно составляют углы 90° (*кубическая сингония* [8]), и что размер тела в периодах решётки одинаков по всем векторам базиса (случай *наночастицы — трёхмерного нанообъекта*). Тогда

$$P_2(x_1, x_2, x_3) \equiv A(x) = \sigma_2 x^2 + \sigma_1 x^1 + \sigma_0, \qquad (4)$$

где $x$ — размер; $\sigma_k$, $k = 1, 2, 3$ — множители.

Очевидно, что числовые значения (3, 4) целочисленны.

Предполагая, что все элементы поверхности одинаковы, и что набор возможных индивидуальных состояний элемента ограничен $n = 0, ..., \eta$, учесть ко-личество сочетаний можно следующим образом. Сумма $\sum \exp(-\beta p \mathrm{v}_\nu)$ выражает статистику нетеплового равновесия объекта и среды (термостата). В свою очередь, пребывание в тепловом равновесии для объекта означает, что для элементов его внешней поверхности можно указать *наиболее вероятное распределение (индивидуальных состояний)* [9, §1-7], соответствующее температуре термостата. То есть, во внешней поверхности в состоянии равновесия в среднем по времени существует (имеет место) формальное распределение $A(x)$ элементов поверхности по $\eta$ квантовым состояниям, то есть, задано множество целых чисел $\{y_0, y_1, ..., y_\eta\}$, элементы которого есть слагаемые указанного разбиения $A(x)$:

$$A(x) = \sum_{j=0}^{\eta} y_j. \tag{5}$$

Комбинаторное осуществление такого разбиения возможно

$$\Pi_C = \prod_{m=0}^{\eta} C\left(\left[\sum_{j=m}^{\eta} y_j\right], y_m\right) \tag{6}$$

способами, где $C(n, k)$ — число сочетаний из $n$ элементов по $k$, т. е., различных подмножеств из $k$ элементов заданного множества из $n$ элементов без учёта порядка выборки [10].

Расчёт величины (6) есть известная комбинаторная задача, решение которой [10]

$$\Pi_C = \left(\sum_{q=0}^{\eta} y_q\right)! \frac{1}{y_0! y_1! ... y_\eta!} = \frac{A(x)!}{y_0! y_1! ... y_\eta!}. \tag{7}$$

Величины $y_q$ рассчитываются, исходя из канонического распределения, как

$$y_q = A(x) \exp(-\beta \varepsilon_m), \tag{8}$$

где $\varepsilon_m$ — энергия, приходящаяся на один элемент и соответствующая пребыванию элемента на уровне $m$; $\beta = (k_B T)^{-1}$ — температурный параметр.

Введя таким образом «действительное» количество сочетаний $\Pi_C$, можно заменить суммирование по статистическим весам $\exp(-\beta p \mathrm{v}_\nu)$ на умножение величины $\Pi_C$ на усреднённый статистический вес:

$$\sum_\nu \exp(-\beta p \mathrm{v}_\nu) = \Pi_C \exp(-\beta p \mathbf{v}), \qquad (9)$$

где $\mathbf{v}$ — величина, имеющая смысл объёма, и по величине соответствующая усреднённому статистическому весу.

Величину $\mathbf{v}$ в начальном приближении можно рассчитать как

$$\mathbf{v} = \mathbf{O}(\alpha^3 x^3), \qquad (10)$$

где $\alpha$ — период кристаллической решётки, принятый для упрощения вида этого выражения одинаковым по всем трём базисным векторам.

Выражения для термодинамических потенциалов содержат натуральный логарифм статистической суммы, то есть, применительно к данному случаю, содержат выражение вида

$$\ln \Pi_C = \ln[A(x)!] - \sum_{q=0}^{\eta} \ln[y_q!]. \qquad (11)$$

Подстановка выражений Стирлинга вместо логарифмов факториалов быстро растущих чисел и последующие очевидные преобразования дают выражение для логарифма числа сочетаний индивидуальных состояний поверхностных элементов

$$\ln \Pi_C = A(x)(\beta U_{1A} + \ln Z_{1A}), \qquad (12)$$

где $Z_{1A} = \sum \exp(-\beta \varepsilon_m)$ — формальное выражение для канонической статистической суммы для одного элемента внешней поверхности; $U_{1A} = Z_{1A}^{-1} \sum \varepsilon_m \exp(-\beta \varepsilon_m)$ — формальное выражение для средней энергии одного элемента внешней поверхности.

Окончательно объёмо-зависимый множитель в логарифме обобщённой по Хиллу статистической суммы имеет вид

$$\ln\left\{\sum_{\nu} \exp(-\beta p \mathbf{v}_\nu)\right\} = \ln \Pi_C + (-\beta p \mathbf{v}) = \qquad (13)$$
$$= A(x)(\beta U_{1A} + \ln Z_{1A}) + P_{1V}$$

где $P_{1V} = -\beta p \mathbf{v}$ — статвес произведения давления в термостате на усреднённый элементарный объём, то есть,

$$P_{1V} = -\mathbf{O}(\beta \cdot p \cdot \alpha^3 x^3). \qquad (14)$$

## Термодинамический потенциал

Поскольку предполагается, что использование выражения (2) каким-то образом уже оправдано (например, с помощью комбинаторно-статистического усреднения [4]: $Z_{1A} = \mathbf{Z}_{1q}$, $U_{1A} = \mathbf{U}_{1q}$), можно перейти, например, к выражению для свободной энергии Гиббса:

$$G = -k_B T \ln \Xi_{1q} = -k_B T \ln\left\{ \Pi_C e^{P_{1V}} \mathbf{Z}_{1q}^{L(x)} \right\}, \qquad (15)$$

где $L(x)$ — форма 3-й степени, выражающая количество элементов в объекте, зависящее от размера и получаемое таким же образом, как и (3, 4).

Подстановка (13) в (15), если дополнительно принять $P_{1V} = -(\beta \cdot p \cdot \alpha^3 x^3)$, даёт

$$G(x, T) = -k_B T \cdot \big(A(x) + L(x)\big) \cdot \ln \mathbf{Z}_{1q} - A(x) \mathbf{U}_{1A} + p \alpha^3 x^3. \qquad (16)$$

В реальных пределах изменения величин, входящих в выражение, второе и третье слагаемые оказываются величинами $\mathbf{o}(10^{-2})$ относительно первого слагаемого. Так что выражение для свободной энергии Гиббса можно рассматривать в упрощённом виде:

$$G(x, T) = -k_B T \big(A(x) + L(x)\big) \ln \mathbf{Z}_{1q}. \qquad (17)$$

Поскольку $A(x) \sim N^{2/3}$ и $L(x) \sim N$ [6, 7], то оказывается, что выражение

(17) совпадает по структуре с выражениями, ранее полученными в самом общем виде для малых объектов различными авторами [1]

$$G = N f + g N^{2/3}, \qquad (18)$$

где $f$ — свободная энергия на одну структурную единицу по всему объёму; $g$ — свободная энергия на структурную единицу, образующую поверхность объекта.

## Размерный множитель

Размер, заданный в безразмерных единицах структуры тела, может быть выражен через количество структурных единиц; для упрощённого случая $x = x_1 = x_2 = x_3$ общий вид такой зависимости «количество вещества-размер» [6, 7]:

$$a(N) = (N \xi - \iota)^{1/3} + \zeta \qquad (19)$$

где $\xi, \iota, \zeta$ — множители, которые, как правило, выражаются рациональными числами и иногда могут быть рассчитаны точно; множитель $\iota$ удовлетворяет условию $\iota > 0$; единицы $a(N)$ — периоды кристаллической решётки.

В этой зависимости неявно содержится также зависимость от габитуса (внешней формы) и структуры тела. Так, для плотнейшей упаковки (гранецентрированной кубической решётки) и идеализированного условно-кубического габитуса [8, 5]

$$a_{ud}(N) = \left[ \frac{N}{4} - \frac{1}{8} \right]^{1/3} - \frac{1}{2}, \qquad (20)$$

для той же упаковки и идеализированного условно-сферического габитуса

$$a_{cф}(N) \simeq 1{,}1 N^{1/3}. \qquad (21)$$

Размерные зависимости количеств элементов в поверхности и объёме для плотнейшей упаковки (гранецентрированной кубической решётки) и идеализированного условно-кубического габитуса [6, 7]

$$L(x) = 28x^3 - 6x^2 + 15x + \mathbf{O}(1), \quad A(x) = 36x^2, \tag{22}$$

где $x$ — размер, который может быть выражен через $x = a(N)$, и тогда

$$L(a(N)) \equiv L(N), \quad A(a(N)) \equiv A(N). \tag{23}$$

Предел отношения

$$\lim_{N \to \infty} \frac{L(N)}{N} \equiv \lim_{N \to \infty} R(N) \tag{24}$$

оказывается по физическому смыслу предельным количеством структурных элементов, приходящимся на одну структурную единицу макроскопического образца. Поскольку $L(N)$ в используемой модели есть степенная функция от размера объекта, функциональная зависимость $R(N)$ имеет смысл *размерного множителя* для количества вещества. Кроме того, эта зависимость напрямую соотносится с *координационным числом* для первой или первой-второй *координационных сфер* [8, с.96–98].

Предел величины размерного множителя есть постоянная для данного типа упаковки:

$$\lim_{N \to \infty} R_{уп}(N) = \mathcal{R}_{уп} = \Theta(K_{ч,уп}/2), \tag{25}$$

где $K_{ч,уп}$ — координационное число, то есть, 14 (6+8) для ПК, 12 для ОЦК и ГЦК, 12 для 20-гранника Маккея и так далее.

## Размерная зависимость теплоёмкости

Один из способов расчёта теплоёмкости — взятие второй производной по температуре от одного из видов свободной энергии:

$$C_p \equiv -T\frac{\partial^2 G}{\partial T^2}\bigg|_p ; \; C_v \equiv -T\frac{\partial^2 F}{\partial T^2}\bigg|_v, \qquad (26)$$

где $C_p$ — теплоёмкость при постоянном давлении; $C_v$ — теплоёмкость при постоянном объёме; $G$ — свободная энергия Гиббса; $F$ — свободная энергия (Гельмгольца); $T$ — температура.

В понятиях используемой модели эти выражения уточняются в виде

$$C \equiv -T \cdot \big(A(N)+L(N)\big) \cdot \frac{\partial^2 \big((-k_B T)\ln Z_{1q}\big)}{\partial T^2}. \qquad (27)$$

Переход к удельным величинам осуществляется по образцу

$$c \equiv \frac{1}{N}C = T \cdot \frac{\big(A(N)+L(N)\big)}{N} \cdot \frac{\partial^2 \big((k_B T)\ln Z_{1q}\big)}{\partial T^2}. \qquad (28)$$

Взятие второй производной в (28) даёт выражение [2] с ожидаемой структурой вида

$$\Phi'' = \frac{1}{k_B T^3}\big(\langle \varepsilon^2 \rangle - \langle \varepsilon \rangle^2\big), \qquad (29)$$

где величина в скобках выражает *квадратичную флуктуацию* величины энергии элемента; $\langle \varepsilon^2 \rangle$ — среднее значение квадрата энергии; $\langle \varepsilon \rangle^2$ — квадрат среднего значения энергии.

Таким образом, выражение удельной теплоёмкости в понятиях данной модели позволяет представить удельную теплоёмкость наноразмерного объекта в виде

$$c \equiv \frac{\big(A(N)+L(N)\big)}{N} \cdot c_{1q}, \qquad (30)$$

где $c_{1q}$ — удельная теплоёмкость, вычисленная для одиночного элемента

(например, одной локальной моды [3]) с учётом поправок на существование элемента в составе наноразмерного объекта [4].

Размерная зависимость удельной теплоёмкости выражается как отношение удельной теплоёмкости наноразмерного объекта к соответствующему макроскопическому значению:

$$R[c](N) \equiv \frac{c(N,T)}{c_0(T)} \simeq$$
$$\simeq \mathcal{R}^{-1}\left(\lambda_3 N_{\xi\iota} + (\lambda_2 + \sigma_2) N_{\xi\iota}^{2/3} + (\lambda_1 + \sigma_1) N_{\xi\iota}^{1/3} + (\lambda_0 + \sigma_0)\right) \quad , \qquad (31)$$

где $\mathcal{R}$ — предельное (макроскопическое) размерного множителя; $N_{\xi\iota} \equiv N\xi - \iota$; $\lambda_k$ — коэффициенты многочлена $L(N)$; $\sigma_k$ — коэффициенты многочлена $A(N)$.

Область определения этой функциональной зависимости формально задаётся как $N = 1 \dots \infty$, но упрощённый — для разработки данной модели — способ построения зависимости линейного размера от $N$ фактически лишает результат смысла при значениях $N$, меньших, чем количество атомов в элементарной ячейке. Это не обессмысливает собственно построенную здесь модель наноразмерного объекта, а лишь накладывает дополнительное ограничение на допустимую мелкость моделируемых объектов.

### Проверка достоверности

Проверку достоверности размерной зависимости теплоёмкости $R[c](N)$ следует осуществлять в отношении образцов, как можно более близких по физическому смыслу к хилловскому коллективу. Из литературы по этому признаку отобраны следующие работы: [11, 12, 13]. Численная проверка соответствия осуществляется обычным образом : 1) экспериментальные данные по теплоёмкости в виде пар «размер—значение» помещаются в массивы данных машинной программы с нормализацией; 2) производится расчёт примерного среднего количества атомов в одиночном объекте (частице), исходя из данных соответствующей работы; 3) Рассчитывается отношение измеренного и макроскопического значений теплоёмкости; 4) нормализованные массивы подвергаются нелинейной подгонке по *методу наименьших квадратов*. Во всех названных случаях получено хорошее совпадение (определяется по величине коэффициента корреляции).

Также полученная модель размерной зависимости проверяется на согласие с другими теоретическими моделями: модифицированной дебаевской моделью Бора (2001) [14] и моделью Прашера—Фелана (1999) [15]. Также получено хорошее совпадение.

## Выводы

Разработанная модель размерной зависимости позволяет с хорошей точностью предсказывать поведение теплоёмкости (и, видимо, иных теплофизических свойств) в наноразмерных объектах, соответствующих по физическому смыслу хилловскому ансамблю (коллективу). Топологическая часть модели требует дальнейшей разработки.

## Список использованных источников

илл.